\documentclass[aps, prl, floats, floatfix, twocolumn , superscriptaddress,noeprint]{revtex4-2}

\usepackage[utf8]{inputenc}
\usepackage{amsfonts,amsmath,amssymb,amsthm,mathtools}
\allowdisplaybreaks
\usepackage{graphicx}
\usepackage[dvipsnames,x11names]{xcolor}

\usepackage{graphicx}
\usepackage{booktabs}
\usepackage{array}
\usepackage{latexsym}
\usepackage{enumitem}
\usepackage[citecolor=MidnightBlue,linkcolor=MidnightBlue,urlcolor=MidnightBlue,colorlinks=true]{hyperref}
\usepackage[title]{appendix}
\usepackage{xspace}
\usepackage[normalem]{ulem}
\usepackage{siunitx}
\usepackage[caption=false]{subfig}
\usepackage{dcolumn}
\usepackage{bm}
\usepackage{mathrsfs}
\usepackage[export]{adjustbox} 
\reversemarginpar

\newcommand{\turnaround}{turnaround\xspace}

\def\i{{i}}

\def\one{\textrm{I}}
\def\two{\textrm{I\!I}}
\def\three{\textrm{I\!I\!I}}
\def\four{\textrm{I\!V}}

\def\region{{\mathcal R}}
\def\regioni{{\region_\i}}
\def\regioniplusone{{\region_{\i+1}}}
\def\regionone{{\region_\one}}
\def\regiontwo{{\region_\two}}
\def\regionthree{{\region_\three}}
\def\regionfour{{\region_\four}}

\def\surface{{\Sigma}}
\def\surfacei{{\surface_\i}}

\def\rad{{a}}
\def\radi{{\rad_\i}}

\def\radone{{\rad_{\one}}}
\def\radtwo{{\rad_{\two}}}
\def\radthree{{\rad_{\three}}}
\def\radfour{{\rad_{\four}}}

\def\lp{\left(}
\def\rp{\right)}

\def\emt{T}

\def\kefe{\varkappa}

\def\eos{EoS\xspace}

\def\gback{g}

\def\cc{\Lambda}

\def\Mass{{M}}

\def\Masstwo{{\Mass_\two}}

\def\Massfour{{\Mass_\four}}

\def\R{{R}}
\def\Rone{{\R_\one}}

\def\Rfour{{\R_\four}}

\def\rta{{r_\textup{TR}}}

\def\rch{{r_\textup{CH}}}
\def\adesitter{{\rad_\textup{dS}}}
\def\anariai{{\rad_\textup{N}}}

\def\den{{E}}%
\def\denc{{\den_c}}

\def\dentwo{{\den_\two}}

\def\prer{{P}}
\def\prerc{{\prer_c}}
\def\prerone{{\prer_\one}}
\def\prertwo{{\prer_\two}}
\def\prerthree{{\prer_\three}}

\def\nuc{{\nu_c}}
\def\nuone{{\nu_\one}}

\def\nuthree{{\nu_\three}}
\def\nufour{{\nu_\four}}

\def\rcrit{r_e}

\theoremstyle{definition}

\def\APJ{Ap. J.}
\def\APJ{Astrophys. J. }
\def\apj{\APJ}

\newcounter{mnotecount}

\newcommand{\mnote}[1]
{\protect{\stepcounter{mnotecount}}$^{\mbox{\footnotesize $\bullet$\themnotecount}}$ 
\marginpar{
\raggedright\tiny
$\!\!\!\!\!\!\,\bullet$\themnotecount: #1} }

\makeatother

\begin{document}
\title{Rigidity and the interpretation of mass with a positive cosmological constant}
\author{Eneko Aranguren}
 \email{eneko.aranguren@ehu.eus}
\author{Raül Vera}%
 \email{raul.vera@ehu.eus}
\affiliation{%
Department of Physics, University of the Basque Country EHU, Bilbao}%
\altaffiliation[Also at ]{EHU Quantum Center, EHU}
          

\begin{abstract}
We provide an explicit counterexample to the rigidity properties
underlying the interpretation of mass in the presence of a positive
cosmological constant $\cc$. Specifically, we construct a family of
regular static stellar configurations satisfying the dominant and
strong energy conditions, and containing no surface layers, for which
the mass parameter of the outer Schwarzschild-de Sitter region can be
positive, zero, or negative.  The zero-mass configuration is precisely the one for which
the outer vacuum
region becomes exactly de Sitter, yielding a spacetime in which a de
Sitter domain coexists with regular perfect-fluid matter. This
contrasts sharply with the $\cc=0$ case, where a Minkowski domain
cannot coexist with perfect-fluid regions satisfying the energy
conditions.
These results show that the static, spherically symmetric realization of the rigidity principle
associated with the positive mass theorem for $\cc=0$ does not carry over to $\cc>0$.
\end{abstract}

\maketitle
The notion of mass in the presence of a positive cosmological constant
$\cc$ remains considerably more subtle than in asymptotically flat
spacetimes (see the review in \cite{Szabados-Tod_2019}). A variety of
notions of mass have been proposed in this setting, associated with
different asymptotic structures, horizons, or quasi-local
constructions. Our concern here, however, is not the choice of a
particular definition of mass, but rather the rigidity properties
from which the physical interpretation of mass derives
in the asymptotically flat case.

For $\cc=0$, the interpretation of mass is rooted in the
positive mass theorem and its rigidity
statement \cite{Schoen_Yau_1979,Witten_1981,Bartnik_1986}:
vanishing mass
implies Minkowski spacetime. The classical existence and uniqueness results for
static perfect-fluid stars \cite{Rendall_1991,A_Carrasco_2007}
provide a particularly transparent
realization of this rigidity principle, as follows.

In static, spherically symmetric spacetimes, the Hawking and Misner-Sharp
masses coincide as functions of the areal radius, and
for $\cc=0$ reduce to the ADM mass at spatial infinity (see e.g. \cite{Hayward:1996}).
The relation between these quasi-local notions of mass and the ADM
mass in the $\cc=0$ case is also reflected in the existence and
uniqueness results for static perfect-fluid stars
\cite{Rendall_1991,A_Carrasco_2007}
through
the matching conditions, that enforce the continuity of the
Misner-Sharp mass across the stellar surface
\cite{Seno:1996_general_matching}.
That fixes the Schwarzschild
mass parameter of the vacuum exterior to be the value of the Misner-Sharp mass at the
boundary. For a perfect fluid, this quantity is the
integral of the energy density over the Euclidean volume element associated with the areal radius,
so positivity follows from the energy conditions.

Crucially, the results of \cite{A_Carrasco_2007} imply that no
additional perfect-fluid matter can be placed outside the stellar boundary.
Consequently, the Schwarzschild exterior extends to spatial
infinity and the Schwarzschild mass parameter coincides with the ADM
mass. Together, these results yield a rigidity property: the
Schwarzschild mass of the configuration vanishes if and only if
the spacetime is Minkowski (thus, there is no star).
Equivalently, the existence of any non-trivial perfect-fluid region
necessarily implies a strictly positive global mass.
In particular, a Minkowski domain cannot coexist with
ordinary perfect-fluid matter (in the absence of surface layers).

A natural question is whether analogous rigidity properties persist for
$\cc>0$. In the same static and spherically symmetric setting,
one would expect that (i) ordinary perfect-fluid regions
satisfying the energy conditions should match to
Schwarzschild-de Sitter (SdS) exteriors with positive mass and,
moreover, that (ii) the existence of a de Sitter domain should force
the entire spacetime to be de Sitter. The model presented here provides
explicit counterexamples to both expectations.

The model consists of a
family of spacetimes containing two concentric perfect fluid regions
--a fluid ball and a fluid (thick) shell-- satisfying the same MIT bag
model equation of state  and separated by a vacuum region.
Fixing the central pressure, the bag constant and $\cc$,
leaves
a one-parameter family of solutions, where the free parameter controls the size of the
intermediate vacuum region.
This vacuum region is
a static SdS domain with positive mass parameter. 
The fluid shell starts at  the innermost boundary with a vanishing pressure,
that attains a maximum and vanishes again at a point where the fluid  matches
a second (outer) vacuum region.

The key point is that the mass parameter of the outer SdS region
varies continuously along the family, taking positive, zero and
negative values.
Therefore, at a critical value of the parameter, 
the outer SdS region
becomes exactly de Sitter. The spacetime then contains a de Sitter
domain with a regular origin coexisting with non-trivial perfect-fluid
regions, all without surface layers.

The model thus provides explicit counterexamples to expectations (i) and (ii).
Expectation (i) fails because the exterior SdS mass can be zero or
negative despite the presence of ordinary perfect-fluid matter
satisfying the strong and dominant energy conditions.
In fact, the negative mass
is unbounded from below.

Statement (ii) fails because a de Sitter domain can coexist with
non-trivial perfect-fluid regions without surface layers. Notably,
this lack of rigidity does not rely on any particular notion of
quasi-local mass.

 Several distinctive features associated with $\cc>0$
have been exploited previously
to construct a variety of non-trivial static matter configurations
(see \cite{Boehmer_eleven:2004,fodor_2008}). In particular, cosmological
and black hole horizons  in SdS permit
models containing two causally disconnected stars located at distinct regular origins and
separated by a vacuum region. In all previously known
constructions, however, the rigidity and positivity properties
described above are preserved: the SdS mass remains positive and
vanishes only in the absence of matter.

The present construction exploits the non-uniqueness of the exterior field for $\cc>0$
\cite{EU_to_be_published},
in sharp contrast with the $\cc=0$
case \cite{A_Carrasco_2007}.
The precise role played by this non-uniqueness in the construction of the model
remains to be understood.

We present next the construction in detail.

\section{Field equations and boundary conditions}

We construct a family of static and spherically symmetric configurations composed of
four concentric, alternating regions of perfect fluid and vacuum,
matched
across symmetry-preserving hypersurfaces
\cite{Seno:1996_general_matching,mps}, with no surface
layers present.  The first region $\regionone$ will contain a center of
symmetry and a perfect fluid content, and will be followed
by an intermediate vacuum region $\regiontwo$, a second
perfect fluid region $\regionthree$, and a final ``outer'' vacuum region $\regionfour$.
  
We consider for each domain a static and spherically symmetric metric written in the form
\begin{equation*}
  \gback = - e^{\nu(r)} dt^2 + dr^2 + \R(r)^2(d\theta^2+\sin^2\theta d\phi^2),
\end{equation*}
where $\R$ is the areal radius function.
We impose
Einstein's Field Equations with
$\cc>0$ for a
perfect fluid energy-momentum tensor
$	\emt_{\alpha\beta} = (\den + \prer)u_\alpha u_\beta + \prer \gback_{\alpha\beta}$,
where $\den$ is the energy density and
$\prer$ the radial pressure of the unit vector field of the fluid
$u=e^{-\nu/2}\partial_{t}$.
The field equations become
\begin{align}
	&\kefe\den=-2\frac{\R^{\prime\prime}}{\R} - \frac{\R^\prime{}^2}{\R^2} + \frac{1}{\R^2} - \cc,\label{EFE_den}\\
	&\kefe\prer= \frac{\R^\prime}{\R}\left(\frac{\R^\prime}{\R}+\nu^\prime\right) - \frac{1}{\R^2} + \cc,\label{EFE_prer}\\
  &\nu''=\kefe \den +\cc+3\frac{\R'^2}{\R^2}-\frac{3}{\R^2}+\frac{\R'}{\R}\nu'-\frac{1}{2}\nu'^2,\label{nu2}
\end{align}
where the prime means derivative with respect to $r$, and $\kefe=8\pi G$
is the gravitational coupling constant (we use $c=1$ throughout).

Using the Misner-Sharp mass, with the vacuum contribution substracted \cite{McVittie_Cahill,Hayward:1996},
\begin{align}
	\Mass:= \frac{\R}{2}\lp1-\R^\prime{}^2\rp - \frac{\cc}{6}\R^3,\label{mass}
\end{align}
equation \eqref{EFE_den}
can be written as
\begin{align}
	\Mass^\prime =&\frac{\kefe}{2}\R^\prime\R^2\den.\label{dM/dr}
\end{align}

A barotropic \eos  $E(P)$
  must be imposed to close the system \eqref{EFE_den}-\eqref{nu2}
  for the three functions $\{\R(r),P(r),\nu(r)\}$.

In particular, the vacuum solution $E=P=0$ 
  is given by either a static domain of SdS, for which
  $\Mass$ is constant and the metric functions satisfy
  \[
    e^\nu=1-\frac{2M}{\R}-\frac{\cc}{3}\R^2,\quad
    \frac{dR(r)}{dr}=\sqrt{1-\frac{2M}{\R}-\frac{\cc}{3}\R^2}
  \]
  for $\R$ in the appropriate range, or by Nariai, for which
  $\R=1/\sqrt{\cc}$ and $e^\nu=\cos^2(\sqrt{\cc} \,r)$.
Observe that de Sitter corresponds to $M=0$.

Regarding the boundary conditions,
regularity
at the origin, $r=0$ in region $\regionone$,
requires that the three functions
admit an expansion
\begin{align}
  &\prerone(r)=\prerc+O(r^2)\\
	&\nuone(r)=\nuc+\frac{\kefe\lp\denc+3\prerc\rp-2\cc}{6} r^2 + O(r^4),\label{expansion_nu_origin}\\
	&\Rone(r)^2 = r^2 -\frac{\kefe\denc+\cc}{9} r^4 + O(r^6),\label{expansion_R_origin}
\end{align}
where
$\denc:=\den(r=0)=\den(\prerc)$,
for some constants $\nuc$ and $\prerc$.
The field equations \eqref{EFE_den}-\eqref{nu2} only depend on $\nu$ through its derivatives,
so $\nuc$ can be chosen freely,
while $\prerc$ is the boundary data
at the origin to be provided.

The boundary conditions for the subsequent regions are
fixed by the matching conditions.
If we denote with a common name $\{t,r,\theta,\phi\}$ the coordinates in each region, chosen so that they run
continuously through the whole spacetime, any
two regions $\regioni$ and $\regioniplusone$, $i=\one,\two,\three$, are joined across
the common boundary $\surfacei:=\{r = \radi\}$
and the matching conditions hold iff
(see e.g. \cite{MRV2,Seno:1996_general_matching})
\begin{align}
	[\nu]_i=0,\quad[\R]_i=0,\quad[\nu^\prime]_i=0,\quad[\R^\prime]_i=0,\label{matching}
\end{align}
where  $[f]_i$ denotes the jump of $f(r)$ on $r=\radi$.
Note that, from \eqref{EFE_prer} and \eqref{mass}, the above conditions imply
that $\prer$ and $\Mass$ are also continuous functions of $r$,
whereas there is no condition for $[\den]_i$,
so $\den(r)$ can be discontinuous.

Observe that if there exists a critical radius $r=\rcrit$ such that $\R^\prime(\rcrit)=0$,
  the spacetime may contain a second origin, which will be regular, or not, depending
  on the behaviour of the metric functions there after integration from $r=0$.

\begin{figure}
	\includegraphics[width=\columnwidth]{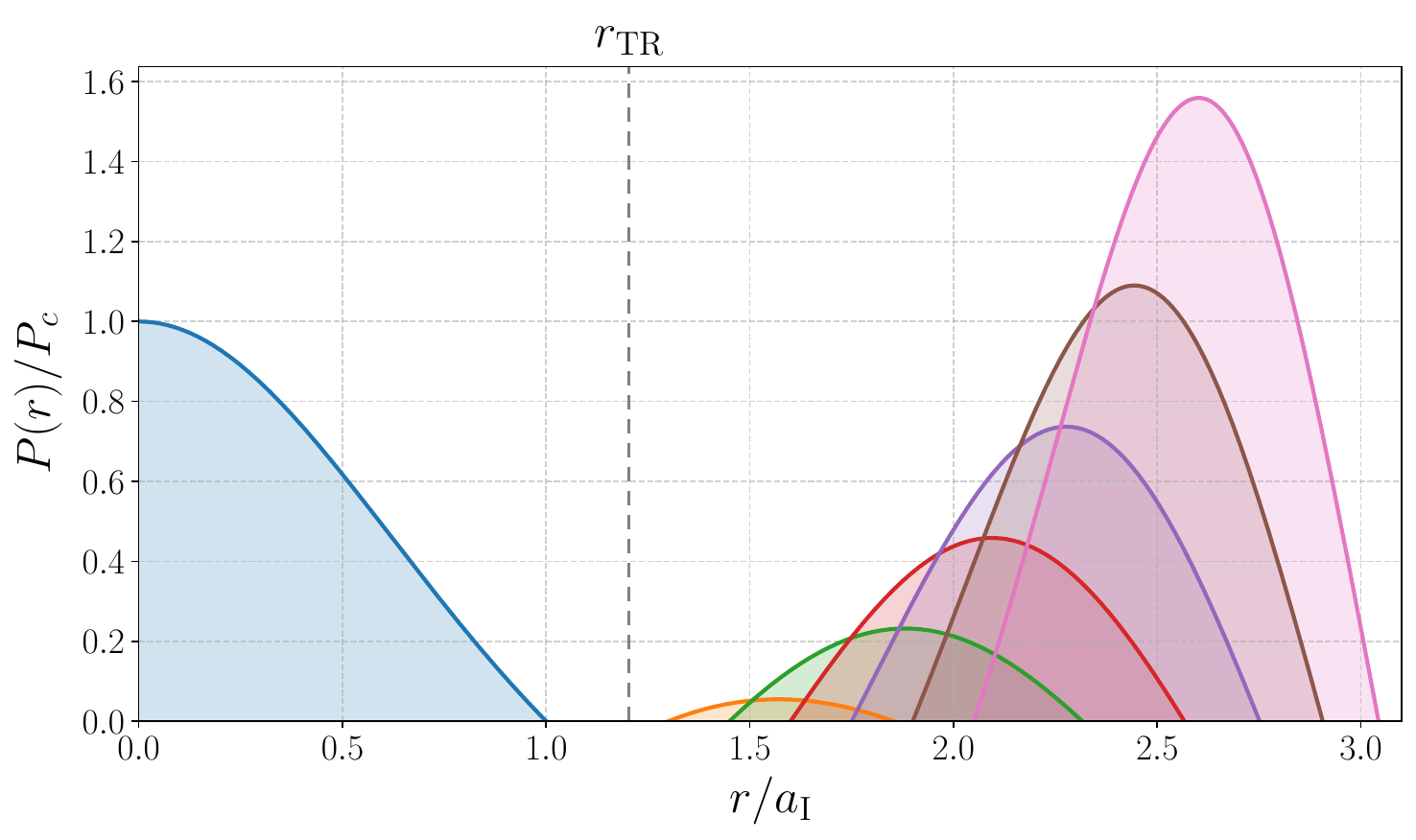}
	\caption{
		Pressure as a function of $r$.
		The blue area represents the central fluid region.
		The dashed line marks the \turnaround radius, $\rta$.
		We show in different colors the fluid shells located at
		$\radtwo/\radone=\{1.3,1.45,1.6,1.75,1.9,2.05\}$.}
	\label{fig:thick_shells}
\end{figure}

\section{The model}
We consider the MIT bag model
(see e.g. \cite{Colpi:1992})
\begin{equation*}
	\den(\prer)=3\prer + 4B,
\end{equation*}
where $B>0$ is the bag constant,
ensuring that $\den(\prer)$ remains strictly positive for all $\prer\geq0$.
In particular, we set $B = 7.44\times10^{-11}$~\si{\m^{-2}}
following \cite{Colpi:1992},
and consider $\prerc=4.92\times10^{-11}$~\si{\m^{-2}}
with $\cc=2.98\times10^{-9}$~\si{\m^{-2}} (we use $G=1$ from now on).

We use these data
in the expansions at the origin  \eqref{expansion_nu_origin}-\eqref{expansion_R_origin}
to integrate numerically the system \eqref{EFE_den}-\eqref{nu2}
from $r=0$ up to $r=\radone$,
where $\radone$ is determined by
$\prerone(\radone)=0$.
Thus, we construct a region $\regionone$ consisting of a perfect fluid ball
satisfying the MIT bag EoS, which can be checked to satisfy, strictly,
the strong and dominant energy conditions,
that matches a vacuum solution in $\regiontwo$.

Thence, we integrate the vacuum region $\regiontwo$
by solving the system
for 
$\prertwo\equiv0$, $\dentwo\equiv0$
with boundary data given by
the matching conditions \eqref{matching}.
Region $\regiontwo$ is
  thus a static region of SdS
  with
  mass $\Masstwo=M(\radone)>0$ and $\cc$.

  There exists a \turnaround radius
$\rta$ in SdS (see e.g. \cite{Mimoso:2009wj}) that marks the point
beyond which a second matter region $\regionthree$
can be sustained \cite{EU_to_be_published}.
This sets the lower bound for the location of the innermost boundary $\radtwo$
for $\regionthree$.
On the other hand, this static domain of SdS
ends at
the cosmological horizon $r=\rch$, and thus
sets the upper bound of $\radtwo$ there.
For the given data, we have $\rta=1.20\radone$ and $\rch=3.50\radone$
\footnote{Here and throughout the manuscript, numerical values are rounded to two decimal places.},
so we take $\radtwo$ as a free parameter
in the range $\radtwo\in(\rta,\rch)$.

For a fixed value of $\radtwo$, we integrate the system \eqref{EFE_den}-\eqref{nu2}
using the same \eos
  starting at $r=\radtwo$ with 
  the boundary conditions \eqref{matching} for $i=\two$.
  For the whole range of $\radtwo$
we obtain that the pressure $\prerthree$ in $\regionthree$ attains a maximum,
which increases with $\radtwo$
[Fig. \ref{fig:thick_shells}],
and vanishes again at some $r=\radthree$,
marking the end of region $\regionthree$.
In all cases $e^{\nuthree(\radthree)}$ is finite.
At $r=\radthree$, as before, $\regionthree$
  matches a vacuum region $\regionfour$,
  which is thus a (second) static vacuum region
  with $\cc$ and mass $\Massfour=M(\radthree)$.

\begin{figure}
	\includegraphics[width=\linewidth]{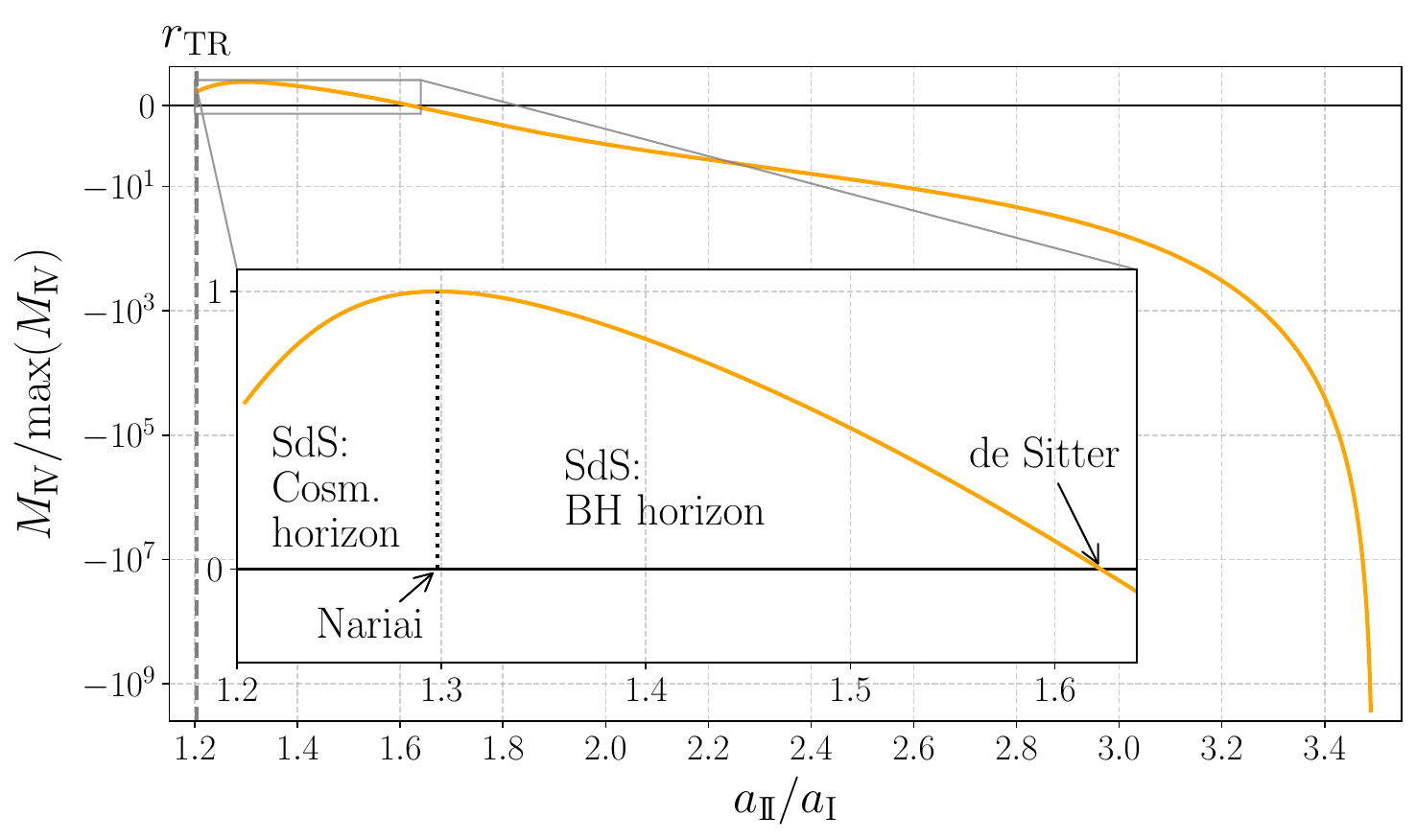}
	\caption{Mass $\Massfour$ of the region $\regionfour$ as a function of $\radtwo$.
          }
	\label{fig:mass-a}
\end{figure}

The crucial fact is that the value
of $\Massfour$ as a function of $\radtwo$
goes continuously from positive to negative values,
with a single zero
at $\radtwo=\adesitter=1.62\radone$ [Fig. \ref{fig:mass-a}].

\begin{figure}
	\includegraphics[width=\columnwidth]{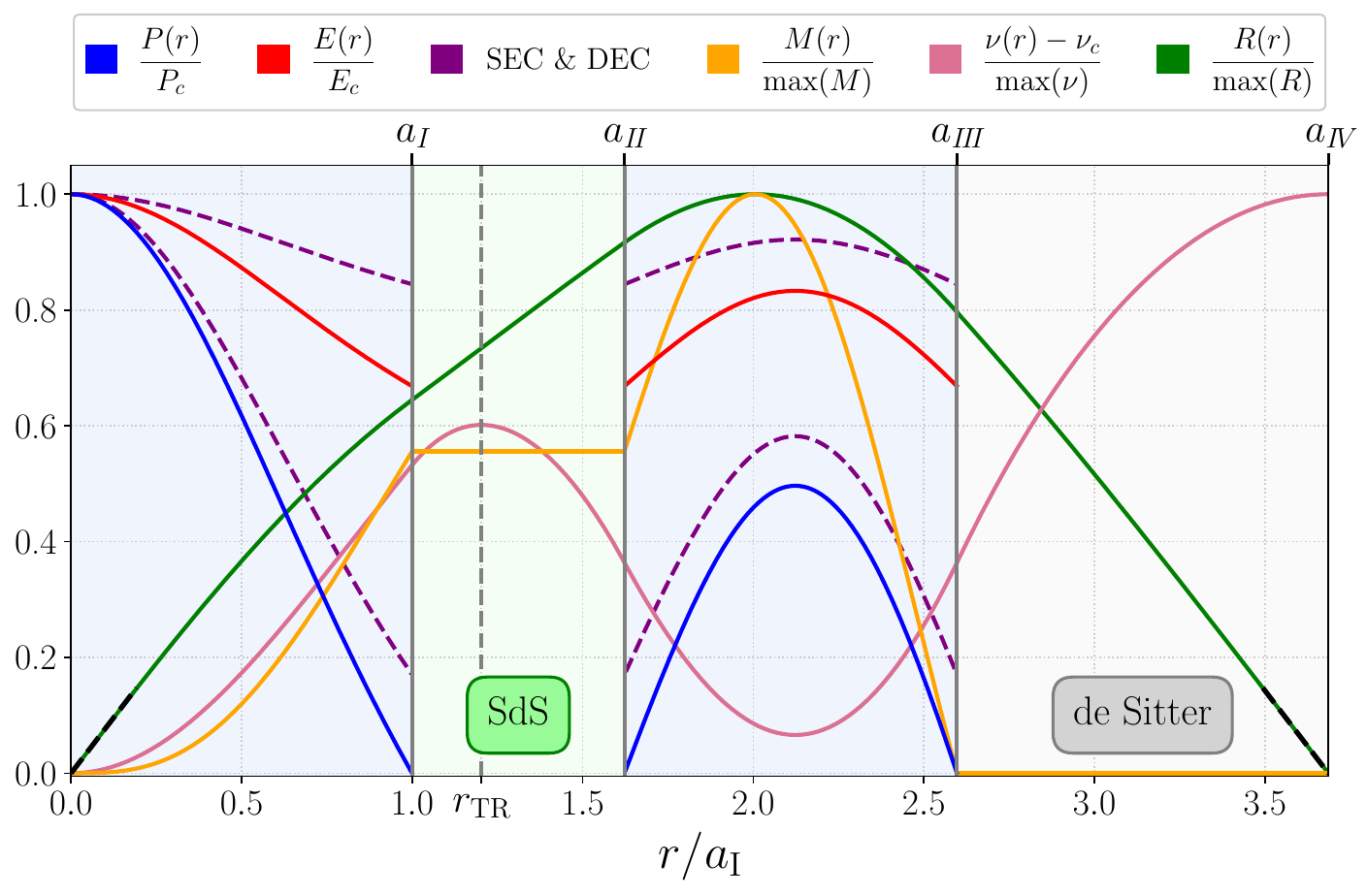}
	\caption{Radial profiles of the relevant quantities
          of the model with a de Sitter region (case $\Massfour=0$).
          The fluid ball $\regionone$
          and the fluid region $\regionthree$
          are shaded in blue.	
          The intermediate vacuum region  $\regiontwo$ is a SdS
          static domain. 
                               $\regionthree$  matches a de Sitter region $\regionfour$
                containing a second origin at $r=3.69\radone$.
               	The curves `SEC' and `DEC' show  $\kefe(\den+\prer)-2\cc$ and $\kefe(\den-\prer)+2\cc$
		normalized by their value at the origin.
		The dashed lines at the origins represent the conditions $\R^\prime = \pm1$.
               }
	\label{fig:plot1}
\end{figure}

Indeed, for $\radtwo<\anariai$, where $\anariai= 1.30\radone$, the pressure vanishes before $\R'$ does, 
so the function $\Mass$ increases with $r$ [c.f. \eqref{dM/dr}]
up to the end of region $\regionthree$, at $r=\radthree$.
However, for $\radtwo\geq \anariai$
there exists in $\regionthree$ a value $r=\rcrit$
at which
$\R$  attains a maximum.
That defines an
equator of the hypersurfaces of constant $t$, and
the timelike hypersurface $r=\rcrit$
constitutes a tube of bifurcating marginal spheres \cite{Hayward:1996}.
More importantly, the maximum of $\R$ also
coincides with the maximum of $\Mass$.
 In the limiting case $\radtwo=\anariai$, both $\R$ and $\Mass$
 attain their maxima at $\rcrit=\radthree$,
 which is, precisely, the end of region $\regionthree$.
For $\radtwo> \anariai$,
the function $\Mass$ decreases from $\rcrit$ onwards until $\regionthree$ ends,
where it gets the value $\Massfour=\Mass(\radthree)$.
It is thus in this range of $\radtwo$ that the function $\Massfour$ can
achieve zero and negative values.
Specifically, we have
$\Massfour>0$ when $\radtwo\in(\rta,\adesitter)$, $\Massfour=0$ at $\radtwo = \adesitter$, and $\Massfour<0$ when $\radtwo>\adesitter$ [Fig. \ref{fig:mass-a}].
Note that $\Massfour$ is unbounded from below.

\begin{figure} 
  \hfill\includegraphics[width=0.43\columnwidth, valign=c]{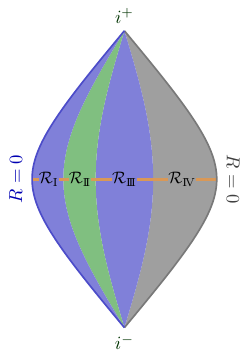}\hfill \includegraphics[width=0.39\columnwidth, valign=c]{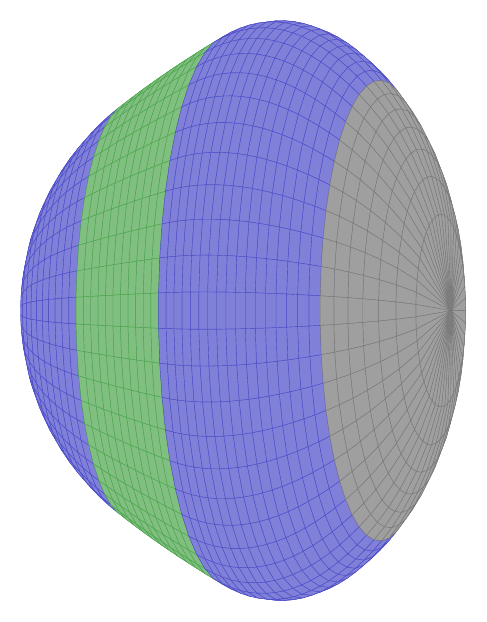}
  \caption{Case $\Massfour=0$. Illustration of the
 conformal diagram of the whole spacetime (left) and a
    surface of constant $t$ and $\theta=\pi/2$ (right).
                Perfect-fluid regions $\regionone$ and $\regionthree$ are colored in blue,
                $\regiontwo$ (SdS) in green, and $\regionfour$ (de Sitter) in grey.}
	\label{fig:dibujo}
\end{figure}

As a result, we obtain the following cases:
\begin{itemize}[noitemsep, parsep=0pt,leftmargin=1em]
\item $\radtwo\in (\rta, \anariai)$:
  $\regionfour$ is a static domain of  SdS with positive mass,
   and $r$ attains a value at which $e^{\nufour(r)}$ vanishes, marking a cosmological horizon.
 \item $\radtwo= \anariai$:
   The matching between $\regionthree$ and $\regionfour$
   is at $r=\rcrit$, thus at $\R'=0$. Consequently, $\regionfour$
   is Nariai.
 \item $\radtwo\in (\anariai,\adesitter)$:
   $\regionfour$ is, again, a static domain of  SdS with positive mass.
   However, $\R'<0$ at the matching,  that is, SdS continues ``inwards''.
   Therefore, $\regionfour$ ends
   when  $r$ reaches a
   black hole horizon.
   \item  $\radtwo=\adesitter$:
     In this case $\Massfour=0$, and therefore  $\regionfour$ becomes a region of (pure) de Sitter.
     There is a second regular origin where $\R$ vanishes again, so that
     each hypersurface
     of constant $t$ is topologically $\mathbb{S}^3$.
We show the relevant functions
in Fig. \ref{fig:plot1}.
Figure \ref{fig:dibujo} illustrates, qualitatively, a
  surface of constant $t$ and $\theta=\pi/2$ isometrically
  embedded in Euclidean $\mathbb{R}^3$,
together with a conformal diagram of the whole spacetime.

\item  $\radtwo\in (\adesitter,\rch)$: $\regionfour$ is a SdS domain
  with negative mass.  There is a value $r=\radfour$ where $\Rfour$
  vanishes, but $\Rfour^\prime(\radfour)<-1$, so that we encounter the usual naked singularity
at the origin
  \cite{Hayward:1996}.

\end{itemize}

\begin{figure}
  \includegraphics[width=0.59\columnwidth,valign=c]{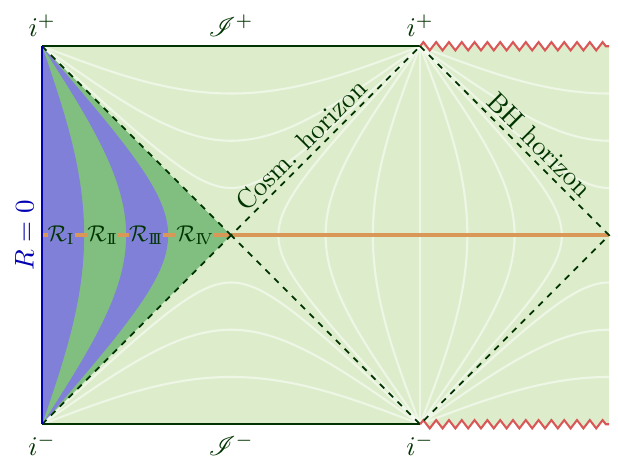}\hfill\includegraphics[width=0.39\columnwidth,valign=c]{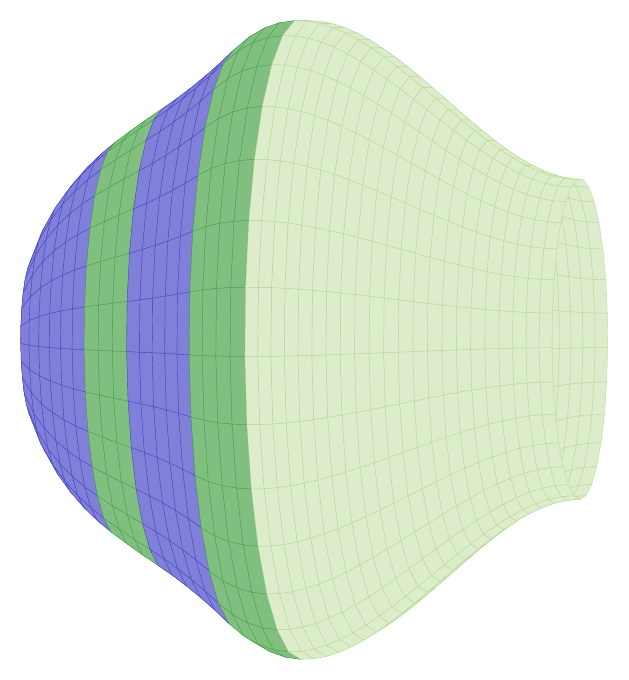}\\
  \includegraphics[width=0.59\columnwidth,valign=c]{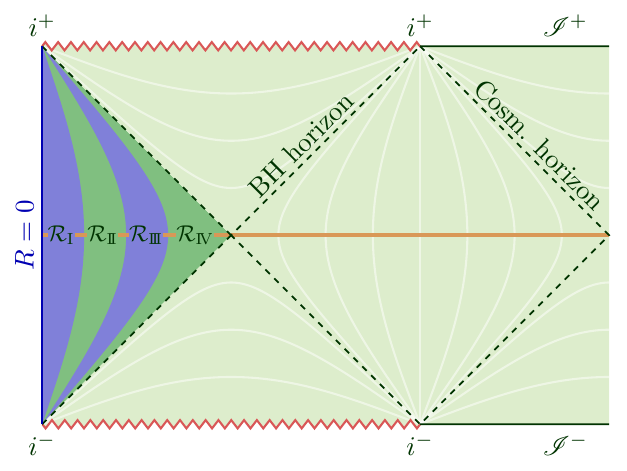}\hfill\includegraphics[width=0.39\columnwidth,angle=0,valign=c]{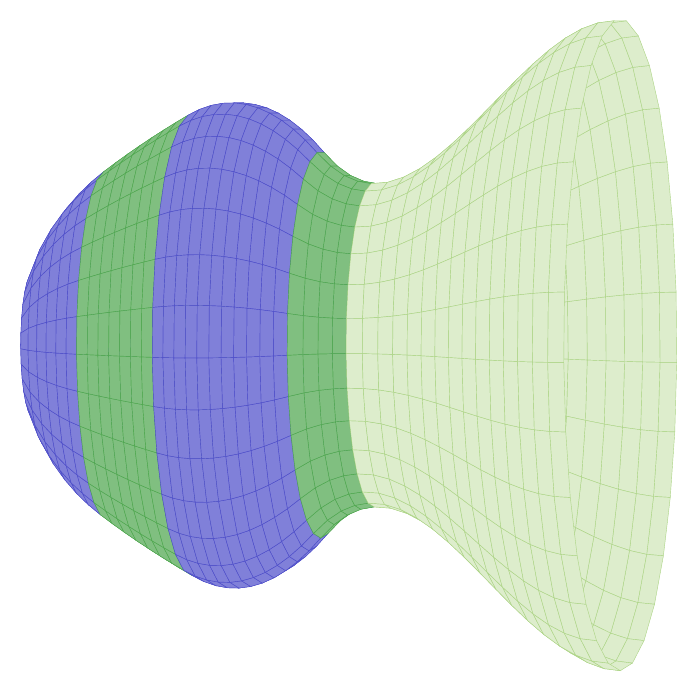}\\
    \hfill\includegraphics[width=0.43\columnwidth,valign=c]{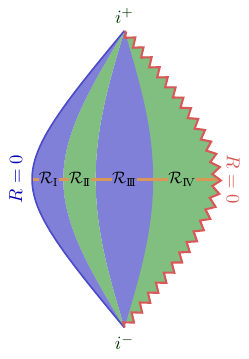}\hfill\includegraphics[width=0.39\columnwidth,angle=0,valign=c]{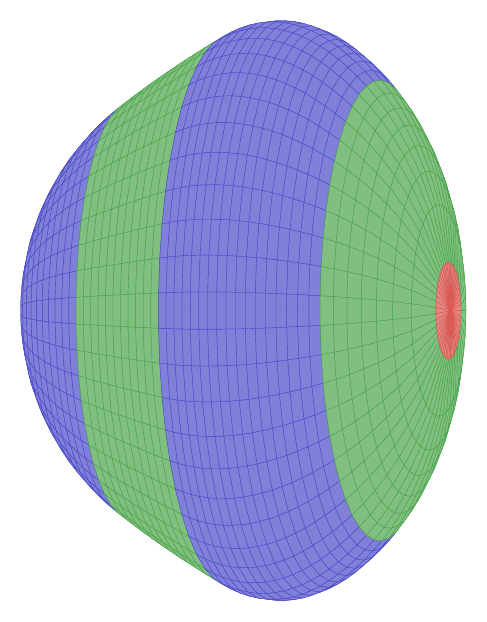}
    \caption{Conformal diagrams and surfaces of constant $t$.
      When $\Massfour>0$,
      $\regionfour$ can be extended to the rest of SdS
      across either a cosmological horizon (top), or a black hole horizon (middle),
      depending on $\radtwo$ (see text).
      In the intermediate situation, $\regionfour$ is Nariai (not shown).
      When $\Massfour=0$ see Fig. \ref{fig:dibujo}.
      Bottom: when $\Massfour<0$, $\regionfour$ contains
      a naked singularity (red).
  }
  \label{fig:extensions}
\end{figure}

\section{Concluding remarks}
Let us note that the present results do not contradict existing
positive mass theorems for asymptotically de Sitter initial data. In
particular, the members of the family with positive SdS mass admit the
usual extensions through their Killing horizons (see Fig. \ref{fig:extensions})
and fall within the
scope of the positivity results of Luo, Xie and Zhang
\cite{LUO201098}, for which
spacelike hypersurfaces with asymptotic flat ends are needed (ending at $i^+$).
By contrast, the critical configuration for
which the outer SdS region becomes exactly de Sitter terminates at a
second regular centre and admits no such asymptotic ends. The
hypotheses of those theorems therefore fail precisely in the case that
realizes the de Sitter domain coexisting with ordinary matter. The
negative-mass members of the family contain instead a timelike naked
singularity, rendering positivity statements inapplicable.

We also briefly note that notions of mass for $\cc>0$ associated with
either asymptotic structure at $\mathscr{I}^{+}$ or isolated
horizons (e.g.
\cite{Ashtekar_Bonga_Kesavan_II,Fernandez-Alvarez_2022,Borghini2020MassII}), as well as known rigidity results for spacetimes asimptotically de Sitter \cite{AnderssonGalloway2002},
can be considered for those members of the family that admit
extensions, namely the positive-mass solutions (see Fig. \ref{fig:extensions}),
but not for the critical zero-mass
configuration, since it possesses neither
asymptotic infinities nor isolated horizons.
It would be
interesting to understand how the present phenomenon relates to notions
of mass for closed universes \cite{Szabados2013}, a
question we leave for future work.

The realization of rigidity in the static, spherically symmetric perfect-fluid setting
  should also be distinguished
  from the classical thin-shell constructions,
  both for $\cc=0$ and $\cc>0$, in which, in particular, vacuum regions
  with different mass parameters are joined across
  a timelike surface layer (see e.g.  \cite{membranes_bonga}),
  and also from configurations joining two de Sitter domains across
  a thin shell  \cite{Forghani2022ClosedUniverse} or
  regions containing anisotropic fluids \cite{natario}, following the idea   of Weyl \cite{Weyl}.
  Here, by contrast, no surface layers are present, and
  the entire construction is supported by regular perfect-fluid regions \footnote{
    A doubly de Sitter configuration is itself contained in the present
      model. It is obtained by setting $\prerc=0$, so that the central fluid ball disappears
    and only the perfect-fluid belt $\regionthree$ remains.}.

The origin of the breakdown of rigidity for $\cc>0$ remains to
be understood.

\begin{acknowledgments}
  We thank Marc Mars and José Natário for a careful reading of the manuscript
  and for fruitful suggestions and discussions.
  Work supported by the Basque Government Grant No. IT1628-22,
  and by Spanish AEI Grant No. 
(funded by MCIN/AEI/10.13039/501100011033 and by “ERDF A way of making Europe”).
E.A. is supported by the Basque Government Grant No. PRE\_2024\_2\_0078.
\end{acknowledgments}

\bibliography{references}

\begin{thebibliography}{28}%
\makeatletter
\providecommand \@ifxundefined [1]{%
 \@ifx{#1\undefined}
}%
\providecommand \@ifnum [1]{%
 \ifnum #1\expandafter \@firstoftwo
 \else \expandafter \@secondoftwo
 \fi
}%
\providecommand \@ifx [1]{%
 \ifx #1\expandafter \@firstoftwo
 \else \expandafter \@secondoftwo
 \fi
}%
\providecommand \natexlab [1]{#1}%
\providecommand \enquote  [1]{``#1''}%
\providecommand \bibnamefont  [1]{#1}%
\providecommand \bibfnamefont [1]{#1}%
\providecommand \citenamefont [1]{#1}%
\providecommand \href@noop [0]{\@secondoftwo}%
\providecommand \href [0]{\begingroup \@sanitize@url \@href}%
\providecommand \@href[1]{\@@startlink{#1}\@@href}%
\providecommand \@@href[1]{\endgroup#1\@@endlink}%
\providecommand \@sanitize@url [0]{\catcode `\\12\catcode `\$12\catcode
  `\&12\catcode `\#12\catcode `\^12\catcode `\_12\catcode `\%12\relax}%
\providecommand \@@startlink[1]{}%
\providecommand \@@endlink[0]{}%
\providecommand \url  [0]{\begingroup\@sanitize@url \@url }%
\providecommand \@url [1]{\endgroup\@href {#1}{\urlprefix }}%
\providecommand \urlprefix  [0]{URL }%
\providecommand \Eprint [0]{\href }%
\providecommand \doibase [0]{https://doi.org/}%
\providecommand \selectlanguage [0]{\@gobble}%
\providecommand \bibinfo  [0]{\@secondoftwo}%
\providecommand \bibfield  [0]{\@secondoftwo}%
\providecommand \translation [1]{[#1]}%
\providecommand \BibitemOpen [0]{}%
\providecommand \bibitemStop [0]{}%
\providecommand \bibitemNoStop [0]{.\EOS\space}%
\providecommand \EOS [0]{\spacefactor3000\relax}%
\providecommand \BibitemShut  [1]{\csname bibitem#1\endcsname}%
\let\auto@bib@innerbib\@empty
\bibitem [{\citenamefont {Szabados}\ and\ \citenamefont
  {Tod}(2019)}]{Szabados-Tod_2019}%
  \BibitemOpen
  \bibfield  {author} {\bibinfo {author} {\bibfnamefont {L.~B.}\ \bibnamefont
  {Szabados}}\ and\ \bibinfo {author} {\bibfnamefont {P.}~\bibnamefont {Tod}},\
  }\bibfield  {title} {\bibinfo {title} {A review of total energy–momenta in
  {GR} with a positive cosmological constant},\ }\href
  {https://doi.org/10.1142/S0218271819300039} {\bibfield  {journal} {\bibinfo
  {journal} {International Journal of Modern Physics D}\ }\textbf {\bibinfo
  {volume} {28}},\ \bibinfo {pages} {1930003} (\bibinfo {year}
  {2019})}\BibitemShut {NoStop}%
\bibitem [{\citenamefont {Schoen}\ and\ \citenamefont
  {Yau}(1979)}]{Schoen_Yau_1979}%
  \BibitemOpen
  \bibfield  {author} {\bibinfo {author} {\bibfnamefont {R.}~\bibnamefont
  {Schoen}}\ and\ \bibinfo {author} {\bibfnamefont {S.-T.}\ \bibnamefont
  {Yau}},\ }\bibfield  {title} {\bibinfo {title} {On the proof of the positive
  mass conjecture in general relativity},\ }\href
  {https://doi.org/10.1007/BF01940959} {\bibfield  {journal} {\bibinfo
  {journal} {Communications in Mathematical Physics}\ }\textbf {\bibinfo
  {volume} {65}},\ \bibinfo {pages} {45} (\bibinfo {year} {1979})}\BibitemShut
  {NoStop}%
\bibitem [{\citenamefont {Witten}(1981)}]{Witten_1981}%
  \BibitemOpen
  \bibfield  {author} {\bibinfo {author} {\bibfnamefont {E.}~\bibnamefont
  {Witten}},\ }\bibfield  {title} {\bibinfo {title} {A new proof of the
  positive energy theorem},\ }\href {https://doi.org/10.1007/BF01208277}
  {\bibfield  {journal} {\bibinfo  {journal} {Communications in Mathematical
  Physics}\ }\textbf {\bibinfo {volume} {80}},\ \bibinfo {pages} {381}
  (\bibinfo {year} {1981})}\BibitemShut {NoStop}%
\bibitem [{\citenamefont {Bartnik}(1986)}]{Bartnik_1986}%
  \BibitemOpen
  \bibfield  {author} {\bibinfo {author} {\bibfnamefont {R.}~\bibnamefont
  {Bartnik}},\ }\bibfield  {title} {\bibinfo {title} {The mass of an
  asymptotically flat manifold},\ }\href
  {https://doi.org/10.1002/cpa.3160390505} {\bibfield  {journal} {\bibinfo
  {journal} {Communications on Pure and Applied Mathematics}\ }\textbf
  {\bibinfo {volume} {39}},\ \bibinfo {pages} {661} (\bibinfo {year}
  {1986})}\BibitemShut {NoStop}%
\bibitem [{\citenamefont {Rendall}\ and\ \citenamefont
  {Schmidt}(1991)}]{Rendall_1991}%
  \BibitemOpen
  \bibfield  {author} {\bibinfo {author} {\bibfnamefont {A.~D.}\ \bibnamefont
  {Rendall}}\ and\ \bibinfo {author} {\bibfnamefont {B.~G.}\ \bibnamefont
  {Schmidt}},\ }\bibfield  {title} {\bibinfo {title} {Existence and properties
  of spherically symmetric static fluid bodies with a given equation of
  state},\ }\href {https://doi.org/10.1088/0264-9381/8/5/022} {\bibfield
  {journal} {\bibinfo  {journal} {Classical and Quantum Gravity}\ }\textbf
  {\bibinfo {volume} {8}},\ \bibinfo {pages} {985} (\bibinfo {year}
  {1991})}\BibitemShut {NoStop}%
\bibitem [{\citenamefont {Carrasco}\ \emph {et~al.}(2007)\citenamefont
  {Carrasco}, \citenamefont {Mars},\ and\ \citenamefont
  {Simon}}]{A_Carrasco_2007}%
  \BibitemOpen
  \bibfield  {author} {\bibinfo {author} {\bibfnamefont {A.}~\bibnamefont
  {Carrasco}}, \bibinfo {author} {\bibfnamefont {M.}~\bibnamefont {Mars}},\
  and\ \bibinfo {author} {\bibfnamefont {W.}~\bibnamefont {Simon}},\ }\bibfield
   {title} {\bibinfo {title} {On perfect fluids and black holes in static
  equilibrium},\ }\href {https://doi.org/10.1088/1742-6596/66/1/012012}
  {\bibfield  {journal} {\bibinfo  {journal} {Journal of Physics: Conference
  Series}\ }\textbf {\bibinfo {volume} {66}},\ \bibinfo {pages} {012012}
  (\bibinfo {year} {2007})}\BibitemShut {NoStop}%
\bibitem [{\citenamefont {Hayward}(1996)}]{Hayward:1996}%
  \BibitemOpen
  \bibfield  {author} {\bibinfo {author} {\bibfnamefont {S.~A.}\ \bibnamefont
  {Hayward}},\ }\bibfield  {title} {\bibinfo {title} {{Gravitational energy in
  spherical symmetry}},\ }\href {https://doi.org/10.1103/PhysRevD.53.1938}
  {\bibfield  {journal} {\bibinfo  {journal} {Phys. Rev. D}\ }\textbf {\bibinfo
  {volume} {53}},\ \bibinfo {pages} {1938} (\bibinfo {year}
  {1996})}\BibitemShut {NoStop}%
\bibitem [{\citenamefont {Fayos}\ \emph {et~al.}(1996)\citenamefont {Fayos},
  \citenamefont {Senovilla},\ and\ \citenamefont
  {Torres}}]{Seno:1996_general_matching}%
  \BibitemOpen
  \bibfield  {author} {\bibinfo {author} {\bibfnamefont {F.}~\bibnamefont
  {Fayos}}, \bibinfo {author} {\bibfnamefont {J.~M.~M.}\ \bibnamefont
  {Senovilla}},\ and\ \bibinfo {author} {\bibfnamefont {R.}~\bibnamefont
  {Torres}},\ }\bibfield  {title} {\bibinfo {title} {General matching of two
  spherically symmetric spacetimes},\ }\href
  {https://doi.org/10.1103/PhysRevD.54.4862} {\bibfield  {journal} {\bibinfo
  {journal} {Phys. Rev. D}\ }\textbf {\bibinfo {volume} {54}},\ \bibinfo
  {pages} {4862} (\bibinfo {year} {1996})}\BibitemShut {NoStop}%
\bibitem [{\citenamefont {Böhmer}(2004)}]{Boehmer_eleven:2004}%
  \BibitemOpen
  \bibfield  {author} {\bibinfo {author} {\bibfnamefont {C.~G.}\ \bibnamefont
  {Böhmer}},\ }\bibfield  {title} {\bibinfo {title} {{Eleven spherically
  symmetric constant density solutions with cosmological constant}},\ }\href
  {https://doi.org/10.1023/B:GERG.0000018088.69051.3b} {\bibfield  {journal}
  {\bibinfo  {journal} {Gen. Rel. Grav.}\ }\textbf {\bibinfo {volume} {36}},\
  \bibinfo {pages} {1039} (\bibinfo {year} {2004})}\BibitemShut {NoStop}%
\bibitem [{\citenamefont {B\"ohmer}\ and\ \citenamefont
  {Fodor}(2008)}]{fodor_2008}%
  \BibitemOpen
  \bibfield  {author} {\bibinfo {author} {\bibfnamefont {C.~G.}\ \bibnamefont
  {B\"ohmer}}\ and\ \bibinfo {author} {\bibfnamefont {G.}~\bibnamefont
  {Fodor}},\ }\bibfield  {title} {\bibinfo {title} {Perfect fluid spheres with
  cosmological constant},\ }\href {https://doi.org/10.1103/PhysRevD.77.064008}
  {\bibfield  {journal} {\bibinfo  {journal} {Phys. Rev. D}\ }\textbf {\bibinfo
  {volume} {77}},\ \bibinfo {pages} {064008} (\bibinfo {year}
  {2008})}\BibitemShut {NoStop}%
\bibitem [{\citenamefont {Aranguren}\ and\ \citenamefont
  {Vera}(2026)}]{EU_to_be_published}%
  \BibitemOpen
  \bibfield  {author} {\bibinfo {author} {\bibfnamefont {E.}~\bibnamefont
  {Aranguren}}\ and\ \bibinfo {author} {\bibfnamefont {R.}~\bibnamefont
  {Vera}},\ }\bibfield  {title} {\bibinfo {title} {{\emph{In preparation}}},\
  }\href@noop {} {\  (\bibinfo {year} {2026})}\BibitemShut {NoStop}%
\bibitem [{\citenamefont {Vera}(2002)}]{mps}%
  \BibitemOpen
  \bibfield  {author} {\bibinfo {author} {\bibfnamefont {R.}~\bibnamefont
  {Vera}},\ }\bibfield  {title} {\bibinfo {title} {Symmetry-preserving
  matchings},\ }\href {https://doi.org/10.1088/0264-9381/19/20/316} {\bibfield
  {journal} {\bibinfo  {journal} {Classical and Quantum Gravity}\ }\textbf
  {\bibinfo {volume} {19}},\ \bibinfo {pages} {5249} (\bibinfo {year}
  {2002})}\BibitemShut {NoStop}%
\bibitem [{\citenamefont {{Cahill}}\ and\ \citenamefont
  {{McVittie}}(1970)}]{McVittie_Cahill}%
  \BibitemOpen
  \bibfield  {author} {\bibinfo {author} {\bibfnamefont {M.~E.}\ \bibnamefont
  {{Cahill}}}\ and\ \bibinfo {author} {\bibfnamefont {G.~C.}\ \bibnamefont
  {{McVittie}}},\ }\bibfield  {title} {\bibinfo {title} {Spherical symmetry and
  mass-energy in general relativity. {I. General }theory},\ }\href
  {https://doi.org/10.1063/1.1665273} {\bibfield  {journal} {\bibinfo
  {journal} {Journal of Mathematical Physics}\ }\textbf {\bibinfo {volume}
  {11}},\ \bibinfo {pages} {1382} (\bibinfo {year} {1970})}\BibitemShut
  {NoStop}%
\bibitem [{\citenamefont {Mars}\ \emph {et~al.}(2022)\citenamefont {Mars},
  \citenamefont {Reina},\ and\ \citenamefont {Vera}}]{MRV2}%
  \BibitemOpen
  \bibfield  {author} {\bibinfo {author} {\bibfnamefont {M.}~\bibnamefont
  {Mars}}, \bibinfo {author} {\bibfnamefont {B.}~\bibnamefont {Reina}},\ and\
  \bibinfo {author} {\bibfnamefont {R.}~\bibnamefont {Vera}},\ }\bibfield
  {title} {\bibinfo {title} {Existence and uniqueness of compact rotating
  configurations in {GR} in second order perturbation theory},\ }\href
  {https://doi.org/10.4310/ATMP.2022.v26.n8.a9} {\bibfield  {journal} {\bibinfo
   {journal} {Advances in Theoretical and Mathematical Physics}\ }\textbf
  {\bibinfo {volume} {26}},\ \bibinfo {pages} {2719} (\bibinfo {year}
  {2022})}\BibitemShut {NoStop}%
\bibitem [{\citenamefont {{Colpi}}\ and\ \citenamefont
  {{Miller}}(1992)}]{Colpi:1992}%
  \BibitemOpen
  \bibfield  {author} {\bibinfo {author} {\bibfnamefont {M.}~\bibnamefont
  {{Colpi}}}\ and\ \bibinfo {author} {\bibfnamefont {J.~C.}\ \bibnamefont
  {{Miller}}},\ }\bibfield  {title} {\bibinfo {title} {{Rotational Properties
  of Strange Stars}},\ }\href {https://doi.org/10.1086/171170} {\bibfield
  {journal} {\bibinfo  {journal} {\apj}\ }\textbf {\bibinfo {volume} {388}},\
  \bibinfo {pages} {513} (\bibinfo {year} {1992})}\BibitemShut {NoStop}%
\bibitem [{\citenamefont {Mimoso}\ \emph {et~al.}(2010)\citenamefont {Mimoso},
  \citenamefont {Le~Delliou},\ and\ \citenamefont {Mena}}]{Mimoso:2009wj}%
  \BibitemOpen
  \bibfield  {author} {\bibinfo {author} {\bibfnamefont {J.~P.}\ \bibnamefont
  {Mimoso}}, \bibinfo {author} {\bibfnamefont {M.}~\bibnamefont {Le~Delliou}},\
  and\ \bibinfo {author} {\bibfnamefont {F.~C.}\ \bibnamefont {Mena}},\
  }\bibfield  {title} {\bibinfo {title} {{Separating expansion from contraction
  in spherically symmetric models with a perfect-fluid: Generalization of the
  Tolman-Oppenheimer-Volkoff condition and application to models with a
  cosmological constant}},\ }\href {https://doi.org/10.1103/PhysRevD.81.123514}
  {\bibfield  {journal} {\bibinfo  {journal} {Phys. Rev. D}\ }\textbf {\bibinfo
  {volume} {81}},\ \bibinfo {pages} {123514} (\bibinfo {year}
  {2010})}\BibitemShut {NoStop}%
\bibitem [{Note1()}]{Note1}%
  \BibitemOpen
  \bibinfo {note} {Here and throughout the manuscript, numerical values are
  rounded to two decimal places.}\BibitemShut {Stop}%
\bibitem [{\citenamefont {Luo}\ \emph {et~al.}(2010)\citenamefont {Luo},
  \citenamefont {Xie},\ and\ \citenamefont {Zhang}}]{LUO201098}%
  \BibitemOpen
  \bibfield  {author} {\bibinfo {author} {\bibfnamefont {M.}~\bibnamefont
  {Luo}}, \bibinfo {author} {\bibfnamefont {N.}~\bibnamefont {Xie}},\ and\
  \bibinfo {author} {\bibfnamefont {X.}~\bibnamefont {Zhang}},\ }\bibfield
  {title} {\bibinfo {title} {Positive mass theorems for asymptotically de
  sitter spacetimes},\ }\href
  {https://doi.org/https://doi.org/10.1016/j.nuclphysb.2009.09.017} {\bibfield
  {journal} {\bibinfo  {journal} {Nuclear Physics B}\ }\textbf {\bibinfo
  {volume} {825}},\ \bibinfo {pages} {98} (\bibinfo {year} {2010})}\BibitemShut
  {NoStop}%
\bibitem [{\citenamefont {Ashtekar}\ \emph {et~al.}(2015)\citenamefont
  {Ashtekar}, \citenamefont {Bonga},\ and\ \citenamefont
  {Kesavan}}]{Ashtekar_Bonga_Kesavan_II}%
  \BibitemOpen
  \bibfield  {author} {\bibinfo {author} {\bibfnamefont {A.}~\bibnamefont
  {Ashtekar}}, \bibinfo {author} {\bibfnamefont {B.}~\bibnamefont {Bonga}},\
  and\ \bibinfo {author} {\bibfnamefont {A.}~\bibnamefont {Kesavan}},\
  }\bibfield  {title} {\bibinfo {title} {Asymptotics with a positive
  cosmological constant. {II. Linear} fields on {de Sitter} spacetime},\ }\href
  {https://doi.org/10.1103/PhysRevD.92.044011} {\bibfield  {journal} {\bibinfo
  {journal} {Phys. Rev. D}\ }\textbf {\bibinfo {volume} {92}},\ \bibinfo
  {pages} {044011} (\bibinfo {year} {2015})}\BibitemShut {NoStop}%
\bibitem [{\citenamefont {Fernández-Álvarez}\ and\ \citenamefont
  {Senovilla}(2022)}]{Fernandez-Alvarez_2022}%
  \BibitemOpen
  \bibfield  {author} {\bibinfo {author} {\bibfnamefont {F.}~\bibnamefont
  {Fernández-Álvarez}}\ and\ \bibinfo {author} {\bibfnamefont {J.~M.~M.}\
  \bibnamefont {Senovilla}},\ }\bibfield  {title} {\bibinfo {title} {Asymptotic
  structure with a positive cosmological constant},\ }\href
  {https://doi.org/10.1088/1361-6382/ac395b} {\bibfield  {journal} {\bibinfo
  {journal} {Classical and Quantum Gravity}\ }\textbf {\bibinfo {volume}
  {39}},\ \bibinfo {pages} {165012} (\bibinfo {year} {2022})}\BibitemShut
  {NoStop}%
\bibitem [{\citenamefont {Borghini}\ and\ \citenamefont
  {Mazzieri}(2020)}]{Borghini2020MassII}%
  \BibitemOpen
  \bibfield  {author} {\bibinfo {author} {\bibfnamefont {S.}~\bibnamefont
  {Borghini}}\ and\ \bibinfo {author} {\bibfnamefont {L.}~\bibnamefont
  {Mazzieri}},\ }\bibfield  {title} {\bibinfo {title} {On the mass of static
  metrics with positive cosmological constant: Ii},\ }\href
  {https://doi.org/10.1007/s00220-020-03739-8} {\bibfield  {journal} {\bibinfo
  {journal} {Communications in Mathematical Physics}\ }\textbf {\bibinfo
  {volume} {377}},\ \bibinfo {pages} {2079} (\bibinfo {year}
  {2020})}\BibitemShut {NoStop}%
\bibitem [{\citenamefont {Andersson}\ and\ \citenamefont
  {Galloway}(2002)}]{AnderssonGalloway2002}%
  \BibitemOpen
  \bibfield  {author} {\bibinfo {author} {\bibfnamefont {L.}~\bibnamefont
  {Andersson}}\ and\ \bibinfo {author} {\bibfnamefont {G.~J.}\ \bibnamefont
  {Galloway}},\ }\bibfield  {title} {\bibinfo {title} {{dS/CFT} and spacetime
  topology},\ }\href {https://doi.org/10.4310/ATMP.2002.v6.n2.a4} {\bibfield
  {journal} {\bibinfo  {journal} {Advances in Theoretical and Mathematical
  Physics}\ }\textbf {\bibinfo {volume} {6}},\ \bibinfo {pages} {307} (\bibinfo
  {year} {2002})}\BibitemShut {NoStop}%
\bibitem [{\citenamefont {Szabados}(2013)}]{Szabados2013}%
  \BibitemOpen
  \bibfield  {author} {\bibinfo {author} {\bibfnamefont {L.~B.}\ \bibnamefont
  {Szabados}},\ }\bibfield  {title} {\bibinfo {title} {On the total mass of
  closed universes},\ }\href {https://doi.org/10.1007/s10714-013-1587-9}
  {\bibfield  {journal} {\bibinfo  {journal} {General Relativity and
  Gravitation}\ }\textbf {\bibinfo {volume} {45}},\ \bibinfo {pages} {2325}
  (\bibinfo {year} {2013})}\BibitemShut {NoStop}%
\bibitem [{\citenamefont {Yang}\ \emph {et~al.}(2023)\citenamefont {Yang},
  \citenamefont {Bonga},\ and\ \citenamefont {Pan}}]{membranes_bonga}%
  \BibitemOpen
  \bibfield  {author} {\bibinfo {author} {\bibfnamefont {H.}~\bibnamefont
  {Yang}}, \bibinfo {author} {\bibfnamefont {B.}~\bibnamefont {Bonga}},\ and\
  \bibinfo {author} {\bibfnamefont {Z.}~\bibnamefont {Pan}},\ }\bibfield
  {title} {\bibinfo {title} {Dynamical instability of self-gravitating
  membranes},\ }\href {https://doi.org/10.1103/PhysRevLett.130.011402}
  {\bibfield  {journal} {\bibinfo  {journal} {Phys. Rev. Lett.}\ }\textbf
  {\bibinfo {volume} {130}},\ \bibinfo {pages} {011402} (\bibinfo {year}
  {2023})}\BibitemShut {NoStop}%
\bibitem [{\citenamefont {Forghani}\ and\ \citenamefont
  {Mazharimousavi}(2022)}]{Forghani2022ClosedUniverse}%
  \BibitemOpen
  \bibfield  {author} {\bibinfo {author} {\bibfnamefont {S.~D.}\ \bibnamefont
  {Forghani}}\ and\ \bibinfo {author} {\bibfnamefont {S.~H.}\ \bibnamefont
  {Mazharimousavi}},\ }\bibfield  {title} {\bibinfo {title} {A closed universe:
  {de Sitter} cosmic gate},\ }\href
  {https://doi.org/10.1016/j.physletb.2022.137411} {\bibfield  {journal}
  {\bibinfo  {journal} {Physics Letters B}\ }\textbf {\bibinfo {volume}
  {834}},\ \bibinfo {pages} {137411} (\bibinfo {year} {2022})}\BibitemShut
  {NoStop}%
\bibitem [{\citenamefont {Natário}()}]{natario}%
  \BibitemOpen
  \bibfield  {author} {\bibinfo {author} {\bibfnamefont {J.}~\bibnamefont
  {Natário}},\ }\href@noop {} {\bibinfo {title} {{Private
  communication}}}\BibitemShut {NoStop}%
\bibitem [{\citenamefont {Weyl}(2013)}]{Weyl}%
  \BibitemOpen
  \bibfield  {author} {\bibinfo {author} {\bibfnamefont {H.}~\bibnamefont
  {Weyl}},\ }\href@noop {} {\emph {\bibinfo {title} {{Space--Time--Matter}}}}\
  (\bibinfo  {publisher} {Project Gutenberg},\ \bibinfo {year}
  {2013})\BibitemShut {NoStop}%
\bibitem [{Note2()}]{Note2}%
  \BibitemOpen
  \bibinfo {note} {A doubly de Sitter configuration is itself contained in the
  present model. It is obtained by setting ${{P}_c}=0$, so that the central
  fluid ball disappears and only the perfect-fluid belt ${{\protect \mathcal
  R}_\protect \textrm {I\protect \!I\protect \!I}}$ remains.}\BibitemShut
  {Stop}%
\end{thebibliography}%

\end{document}